\documentclass[twocolumn,amsmath,amssymb,floatfix,superscriptaddress]{revtex4}

\usepackage{graphicx}

\newcommand{\id}{\openone}

\begin{document}
\title{Multi-particle entanglement manipulation under
positive partial transpose preserving operations}
\author{Satoshi Ishizaka}
\affiliation{Fundamental and Environmental Research Laboratories,
NEC Corporation, 34 Miyukigaoka, Tsukuba, 305-8501, Japan}
\affiliation{PRESTO, Japan Science and Technology Agency, 4-1-8
Honcho Kawaguchi, 332-0012, Japan}
\author{Martin B.\ Plenio}
\affiliation{QOLS, Blackett Laboratory, Imperial College London,
Prince Consort Road, London SW7 2BW, UK} \affiliation{Institute
for Mathematical Sciences, Imperial College London, 53 Exhibition
Road, London SW7 2BW, UK}
\date{\today}
\begin{abstract}
We consider the transformation of  multi-partite  states in the
single copy setting under positive-partial-transpose-preserving
operations (PPT-operations)
 and obtain  both qualitative and quantitative
results.  Firstly, for some pure state transformations that are
impossible under local operations and classical communication
(LOCC), we demonstrate that they become possible with a
surprisingly large success probability under PPT-operations.
Furthermore, we clarify  the convertibility of arbitrary
multipartite pure states under PPT-operations, and show that a
drastic simplification in the classification of pure state
entanglement occurs when the set of operations is switched from
LOCC to PPT-operations. Indeed, the infinitely many types of
LOCC-incomparable entanglement are reduced to only one type under
the action of PPT-operations. This is a clear manifestation of the
increased power afforded by the use of PPT-bound entanglement.
In addition, we further enlarge the set of operations
to clarify the effect of another type of bound
entanglement, multipartite unlockable bound entanglement, and show
that a further simplification occurs. As compared to pure states a
more complicated situation emerges in the mixed state settings.
While single copy distillation becomes possible under
PPT-operations for some mixed states  it remains impossible for
other mixed states.
\end{abstract}
\maketitle
\section{Introduction}
\label{sec: Introduction}
Constraints and resources are intimately related in physics. If we
impose a constraint on a physical setting then certain tasks
become impossible. A resource must be made available to overcome
the restrictions imposed by the constraints. By definition such a
resource cannot be created employing only the constrained set of
operations but it may be manipulated and transformed under these
operations. That the amount of resource does not increase under
any operation satisfying the constraint emerges then as a
fundamental law, for example in entanglement theory
\cite{Plenio98a,Eisert03a}.

One example of particular importance is the restriction to local
quantum operations and classical communication (LOCC). The
resource that is implied by this constraint are non-separable
states and in particular pure entangled states such as singlet
states, neither of which can be created by LOCC alone. This
setting gives rise to a theory of entanglement as a resource under
LOCC operations.

Any such theory of entanglement as a resource will generally aim
to  provide mathematical structures to allow answers to  three
questions, namely (1) the characterization of entanglement, (2)
the manipulation of entanglement and (3) the quantification of the
entanglement resource \cite{Plenio98a,Eisert03a} under the given
constraint. Of particular interest is the question  of how many
inequivalent types of entanglement  exist  within such a theory.
In the limit of infinitely many identically prepared copies of
bipartite pure states, entanglement can be inter-converted
reversibly \cite{Bennett96b} and it is reasonable to say that
there is only one type of pure bipartite entanglement. Even for
pure states, the situation changes dramatically when we consider
the single copy setting. It has been shown that the Schmidt rank
of bipartite pure states cannot be increased by LOCC
\cite{Lo97a,Nielsen99a,Vidal99a,Jonathan99a,Jonathan99b}. At the
single copy level, the convertibility of bipartite entanglement is
then characterized by the Schmidt-rank \cite{Dur00b}. For finite
dimensional systems a state can be converted to another one with
finite probability exactly if the Schmidt-number of the target
state is not larger than that of the initial state. In  a
tripartite setting the situation is more complicated. Here  it is
well-known for example that a GHZ state cannot be transformed to a
$W$ state and vice versa \cite{Dur00b}. These  states are then
said to be incomparable.  It can be shown that there are two
incomparable types of tripartite entanglement in three qubits
systems. The situation is  even more complicated in multipartite
settings composed by many parties
\cite{Verstraete02a,Miyake03a,Verstraete03a,Briand03a,Miyake04a}
or infinite dimensional bipartite systems \cite{Owari04a,Eisert
SP02}, where there are many (possibly infinitely many)
incomparable types of entanglement.

A different setting is presented by the concept of partial time
reversal or partial transposition \cite{Peres96a}. For two qubits,
states that remain positive under partial transposition (denoted
as PPT-states) are exactly the separable states
\cite{Horodecki96a} but for higher dimensions this is generally
 not the case as there are PPT-states that are inseparable
\cite{Horodecki98a}. This motivates the definition of the set of
positive-partial-transpose-preserving operations (PPT-operations),
defined as operations that map any PPT-state into another
PPT-state \cite{Rains01a}. In this case, the resource are states
that are not PPT (denoted as NPT-states). In the  single copy
setting  for pure states, it has been shown that both under
PPT-operations \cite{Audenaert03a} and with LOCC supported by
PPT-bound entanglement \cite{Ishizaka04b} the Schmidt-number can
be increased  so that state transformations become possible that
are strictly impossible under LOCC.  Furthermore, there are mixed
state transformations that are reversible in the asymptotic
setting \cite{Audenaert03a}.  This suggests that a theory of
entanglement under PPT-operations might have a much simpler
structure than that under the LOCC constraint.

In this paper, we focus  attention  on the entanglement
manipulation  under PPT operations  in the non-asymptotic, single
copy setting  to explore what simplification occur. We
  consider  PPT state transformation in
multipartite settings and obtain both qualitative and quantitative
results. In Sec.\ \ref{sec: Basic notation}, the general settings
 and notations  of PPT preserving operations are
introduced. In Secs.\ \ref{sec: Bipartite} and \ref{sec:
Tripartite}, we first demonstrate that the transformations of pure
states that are impossible under LOCC become possible with a
surprisingly  large success probability  when  employing trace
preserving PPT-operations. In Secs.\ \ref{sec: Trace
non-preserving}, a rather tractable scheme of trace non-preserving
PPT-operations is introduced and discussed. We will then
completely clarify the convertibility of all multipartite pure
states under PPT-operations in Sec.\ \ref{sec: Convertibility of
pure states}. In Sec.\ \ref{sec: Unlockable states}
we enlarge the set of operations beyond that of PPT-operations
to consider the effect of multipartite unlockable
bound entangled states. In Sec.\ \ref{sec: Single copy
distillation}, we will consider the transformation of a single
copy of mixed states into pure entangled states, i.e. the single
copy distillation under PPT-operations. A summary and conclusion
is given in Sec.\ \ref{sec: Summary}.
\section{Basic notation}
\label{sec: Basic notation}
To begin with, let us denote ${\cal H}(V)$ (${\cal H}(V')$) the
space of Hermitian operators on the Hilbert space $V$ ($V'$). A
superoperator $\Psi$ from $V$ to $V'$ is a linear transformation
from ${\cal H}(V)$ to ${\cal H}(V')$. There is a natural
isomorphism  \cite{Rains01a} which associates with superoperators
$\Psi:{\cal H}(V) \rightarrow {\cal H}(V')$ a Hermitian operator
$\Omega(\Psi)\in{\cal H}(V) \otimes {\cal H}(V')$ such that for
all $A\in{\cal H}(V)$ and $B\in{\cal H}(V')$ we have
\begin{equation}
    \hbox{tr}\{ \Psi(A)B\}= \hbox{tr}\{\Omega(\Psi) A\otimes B\}.
    \label{iso}
\end{equation}
Maps that are trace non-increasing then satisfy
\begin{equation}
    \hbox{tr}_{V'}\{\Omega({\Psi})\} \le \id_{V}
\end{equation}
with equality if $\Psi$ is trace preserving.  A superoperator
$\Psi$ is  called  positive if for any $A\ge 0$ we have
$\Psi(A)\ge 0$ and it is called completely positive if
$\id_W\otimes \Psi\ge 0$ for any space $W$. Following
\cite{Rains01a} complete positivity (CP) of $\Psi$ can be verified
by checking
\begin{equation}
    \Omega(\Psi)^{\Gamma_V}\ge 0
\end{equation}
where $\Gamma_V$ denotes the partial transposition with respect to $V$.

An additional concept comes into play when we consider multipartite systems.
A CP-map on bipartite systems
$\Psi:
{\cal H}(V_A)\otimes {\cal H}(V_B)\rightarrow
{\cal H}(V'_A)\otimes {\cal H}(V'_B)$
is called positive partial transpose preserving (PPT) \cite{Rains01a}, if we
have $\Gamma_A\circ\Psi\circ\Gamma_A\!\ge\!0$
($\Gamma_B\circ\Psi\circ\Gamma_B\!\ge\!0$) for the partial transposition map
$\Gamma_A$ ($\Gamma_B$) with respect to party $A$ ($B$).
On the level of the state $\Omega(\Psi)$, this condition reads
\[
    (\Omega(\Psi)^{\Gamma_V})^{\Gamma_{V_A}\otimes\Gamma_{V'_A}} \ge 0
    \hbox{~~or~~}
    (\Omega(\Psi)^{\Gamma_V})^{\Gamma_{V_B}\otimes\Gamma_{V'_B}} \ge 0
\]
 where $\Gamma_{V_A}$ ($\Gamma_{V'_A}$) denotes partial
transposition applied to space $V_A$ $(V'_A)$. In the bipartite
case, there are two  equivalent  choices for the partial
transposition. In the tripartite setting however, there are three
different possible partial transpositions that are generally {\em
not} equivalent. A CP map $\Psi: {\cal H}(V_A)\otimes {\cal
H}(V_B)\otimes {\cal H}(V_C)\rightarrow {\cal H}(V'_A)\otimes
{\cal H}(V'_B)\otimes {\cal H}(V'_C)$  will be called PPT in the
following  if
\begin{equation}
    (\Omega(\Psi)^{\Gamma_V})^{\Gamma_{V_i}\otimes\Gamma_{V'_i}} \ge 0
    \label{eq: PPT conditions}
\end{equation}
for all $i\!=\!A$, $B$ and $C$.

Let us now consider the transformation of a state $\rho\in{\cal
H}(V)$ into a state $\sigma\in{\cal H}(V')$ with the probability
of $p(\rho\!\rightarrow\!\sigma)$. For this probabilistic
transformation, we construct the trace preserving CP-PPT map with
two outcomes, one that gives $\sigma$ and one that gives some
other state. The two parts are given by CP-PPT maps $\Psi$ and
$\psi$, respectively. The  associated Hermitian operators are
denoted by  $\Omega$ and $\omega$. The map $\Psi$ then satisfies
$\Psi(\rho)\!=\!p(\rho\!\rightarrow\!\sigma)\sigma$ or
\begin{equation}
\hbox{tr}\{\Psi(\rho)(\id-\sigma)\}=
\hbox{tr}\{\Omega(\Psi)\rho\otimes(\id-\sigma)\}=0
\end{equation}
when $\sigma$ is a pure state. The success probability is then
given by
\begin{eqnarray*}
p(\rho\rightarrow\sigma)=\hbox{tr}\{\Psi(\rho)\}
&=&\hbox{tr}\{\Omega(\Psi)\rho\otimes\id\} \\
&=&\hbox{tr}\{\Omega(\Psi)\rho\otimes\sigma\}.
\end{eqnarray*}
 The PPT-map $\psi$, on the other hand, does not suffer
any constraint other than the condition  of trace-preservation for
$\Psi+\psi$. On the level of states, the trace-preserving
condition is
\begin{equation}
    \hbox{tr}_{V'}\{\Omega+\omega\} = \id_V,
    \label{eq: trace-preserving}
\end{equation}
where, as we will do in the remainder of this paper, we have
dropped the $\Psi$ ($\psi$) in $\Omega(\Psi)$ ($\omega(\psi)$) for
brevity. It should be noted that a rather simple structure can be
assumed for $\omega$ without loss of generality. Let us consider a
map $\chi$ which maps arbitrary states in ${\cal H}(V)$ into a
maximally mixed state of $\id_{V'}/\hbox{dim}\{{\cal
H}(V')\}\!\in\!{\cal H}(V')$. This map is a trace-preserving
CP-PPT map since the corresponding state is
$\id_{V}\otimes\id_{V'}/\hbox{dim}\{{\cal H}(V)\otimes{\cal
H}(V')\}$. Therefore, a composed map of $\chi\circ\psi$ is a
CP-PPT map if $\psi$ is a CP-PPT map. Furthermore, if
$\Psi\!+\!\psi$ is trace preserving, $\Psi\!+\!\chi\circ\psi$ is
also trace preserving, and hence the replacement of $\psi$ by
$\chi\circ\psi$ does not alter $p(\rho\!\rightarrow\!\sigma)$. One
may then assume $\psi\!=\chi\circ\psi$ since the output of $\psi$
is arbitrary. On the level of the state, this assumption is
\begin{equation}
    \omega=\omega_V\otimes \frac{\id_{V'}}{\hbox{dim}\{{\cal H}(V')\}}.
\end{equation}
In the subsequent Secs.\ \ref{sec: Bipartite} and \ref{sec:
Tripartite}, we maximize $p(\rho\!\rightarrow\!\sigma)$ for some
important classes of pure states in both bipartite and tripartite
settings.  In particular, we demonstrate that transformations of
pure states that are impossible under LOCC can be achieved under
PPT operations with a surprisingly large success probability.
\section{Conversion of maximally entangled states}
\label{sec: Bipartite}
 For two d-dimensional systems we denote the maximally
entangled state by
$P^+_d\!\equiv\!|\phi^+_d\rangle\langle\phi^+_d|$ where
\begin{displaymath}
    |\phi^+_d\rangle=\frac{1}{\sqrt{d}}\sum_{i=0}^{d-1}|ii\rangle.
\end{displaymath}
 In the single copy setting, it is known that LOCC cannot
increase the Schmidt rank of a pure state
\cite{Lo97a,Nielsen99a,Vidal99a,Jonathan99a,Jonathan99b}.
Therefore,  $p(P^+_d\!\rightarrow\!P^+_{d'})=0$ for LOCC
transformation whenever  $d'\!>\!d$.

 In the following we proceed with the construction of the
CP-PPT maps $\Psi$ and $\psi$ that maximize the success
probability  for this transformation.  For $d'\!>\!d$ this amounts
to the maximization of
\begin{equation}
    p(P^+_d \rightarrow P^+_{d'})=
    \hbox{tr}\{\Omega P^+_d\otimes P^+_{d'}\}
\end{equation}
under the constraints
\begin{eqnarray}
    \hbox{tr}\{\Omega P^+_d \otimes(\id-P^+_{d'})\} &=& 0, \;\;
    \hbox{tr}_{V'}\{\Omega+\omega\} = \id, \nonumber\\
    (\Omega^{\Gamma_{V_A}\otimes\Gamma'_{V_A}})^{\Gamma_{V}} &\ge& 0,
    \hspace*{1.45cm}\Omega^{\Gamma_{V}} \ge 0,      \label{eq: bipartite constraints}\\
    (\omega^{\Gamma_{V_A}\otimes\Gamma'_{V_A}})^{\Gamma_{V}} &\ge& 0,
    \hspace*{1.45cm} \omega^{\Gamma_{V}} \ge 0,\nonumber
\end{eqnarray}
where $P^+_d\!\in\!{\cal H}(V)$ and $P^+_{d'}\!\in\!{\cal H}(V')$.
Since both $P^+_d\!\otimes\!P^+_{d'}$ and
$P^+_d\!\otimes\!(\id\!-\!P^+_{d'})$ are invariant under the local
unitary transformation of $U_1 \otimes U^*_1 \otimes U_{2}
\otimes U^*_{2}$ with $U_1$ and $U_{2}$ being arbitrary unitary
operators, it suffices to consider $\Omega$ and $\omega$  that are
 invariant under these local operations, i.e.
\begin{eqnarray*}
    \Omega&=& a_1 P^+_d \otimes P^+_{d'}
    + a_2 (\id-P^+_d) \otimes P^+_{d'} \\
    &+& a_3 P^+_d \otimes \frac{\id-P^+_{d'}}{{d'}^2-1}
    + a_4 (\id-P^+_d) \otimes \frac{\id-P^+_{d'}}{{d'}^2-1}, \\
    \omega_V &=& b_1 P^+_d  + b_2 (\id-P^+_d).
\end{eqnarray*}
 The first two constraints in Eq.\ (\ref{eq: bipartite
constraints}) yield  $a_3\!=\!0$, $b_1\!=\!1-a_1$, and
$b_2\!=\!1-a_2-a_4$.  These equalities can be used to eliminate
$b_1$ and $b_2$ in the remaining constraints. The remaining
constraints then result in
\begin{eqnarray*}
    1\ge a_1 \ge 0, \;\; a_2 \ge 0, \;\; a_4 \ge 0, \;\;
    1\ge a_2 + a_4, &&  \\
    (d'+1)a_1+(d'+1)(d-1)a_2+(d-1) a_4&\ge& 0, \\
    -(d'+1)a_1+(d'+1)(d+1)a_2+(d+1)a_4&\ge& 0, \\
    -(d'-1)a_1-(d'-1)(d-1)a_2+(d-1)a_4&\ge& 0, \\
    (d'-1)a_1-(d'-1)(d+1)a_2+(d+1) a_4&\ge& 0, \\
    -a_1-(d-1)a_2-(d-1)a_4+d &\ge& 0, \\
    a_1-(d+1)a_2-(d+1)a_4+d &\ge& 0.
\end{eqnarray*}
 The constraints in the first row arise from
$\omega^{\Gamma_{V}}\!\ge\!0$ and $\Omega^{\Gamma_{V}}\!\ge\!0$. The
last two rows are due to
$(\omega^{\Gamma_{V_A}\otimes\Gamma'_{V_A}})^{\Gamma_{V}}\!\ge\!0$
and the remaining for inequalities arise from
$(\Omega^{\Gamma_{V_A}\otimes\Gamma'_{V_A}})^{\Gamma_{V}}\!\ge\!0$.
The maximization of $p(P^+_d\!\rightarrow\!P^+_{d'})\!=\!a_1$
under these constraints is a linear program and we can identify
the optimal solution as $a_1\!=\!d(d\!-\!1)/(dd'\!+\!d'\!-\!2d)$,
$a_2\!=\!0$, and $a_4\!=\!d(d'\!-\!1)/(dd'\!+\!d'\!-\!2d)$.
Consequently,  for $d'>d$ the optimal probability for the
transformation of $P^+_{d}$ into $P^+_{d'}$, thereby increasing
the Schmidt rank,  under PPT-operations is  given by
\begin{equation}
    p(P^+_d \rightarrow P^+_{d'})= \frac{d(d-1)}{dd'+d'-2d}.
    \label{eq: probability of mes}
\end{equation}
 We emphasize that this success probability is nonzero
even when $d'\!>\!d\!\ge\!2$, while it is strictly zero for the
LOCC transformation.
\section{Conversion from GHZ to $W$ state}
\label{sec: Tripartite}
In the tripartite setting, it is well-known that the success
probability  $p(GHZ\!\rightarrow\!W)=0$ for the LOCC
transformation from a single copy of
\begin{displaymath}
    |GHZ\rangle = \frac{|000\rangle + |111\rangle}{\sqrt{2}}
\end{displaymath}
to
\begin{displaymath}
    |W\rangle = \frac{|001\rangle + |010\rangle +
    |100\rangle}{\sqrt{3}}
\end{displaymath}
 \cite{Dur00b}. In the following we will demonstrate that this
is not the case when we consider PPT-operations. To this end, we
maximize
\begin{equation}
    p(\rho_{GHZ}\rightarrow\rho_W)=
    \hbox{tr}\{\Omega\rho_{GHZ}\otimes\rho_W\}
    \label{optcom}
\end{equation}
under the constraints for $i=A,B,C$,
\begin{eqnarray*}
    \hbox{tr}\{\Omega\rho_{GHZ}\otimes(\id-\rho_{W})\} &=& 0, \;\;
    \hbox{tr}_{V'}\{\Omega+\omega\} = \id, \nonumber \\
    (\Omega^{\Gamma_{V_i}\otimes\Gamma'_{V_i}})^{\Gamma_{V}} &\ge& 0,
    \hspace*{1.45cm} \Omega^{\Gamma_{V}} \ge 0, \;\;\nonumber\\
    (\omega^{\Gamma_{V_i}\otimes\Gamma'_{V_i}})^{\Gamma_{V}} &\ge& 0,
    \hspace*{1.45cm} \omega^{\Gamma_{V}} \ge 0, \;\;
\end{eqnarray*}
 where $\rho_{GHZ}\!=\!|GHZ\rangle\langle GHZ|\!\in\!{\cal
H}(V)$ and $\rho_W\!=\!|W\rangle\langle W|\!\in\!{\cal H}(V')$.

The solution of the problem is greatly aided by  the use of  a
number of symmetries. Indeed, both  the states
$\rho_{GHZ}\!\otimes\!(\id\!-\!\rho_{W})$ and
$\rho_{GHZ}\!\otimes\!\rho_{W}$ are invariant under the local
operations
\begin{eqnarray*}
    &(a)& X \otimes X\otimes X\otimes \,\id\,\otimes\,\id\,\otimes\,\id\,,\\
    &(b)& Z\,\otimes Z\,\otimes \,\id\, \otimes \,\id\,\otimes\,\id\,\otimes\,\id\,,\\
    &(c)& \,\id\,\otimes Z\,\otimes Z\,\otimes \,\id\,\otimes\,\id\,\otimes\,\id\,,\\
    &(d)& \,\id\,\otimes\,\id\,\otimes\,\id\,\otimes Z\,\otimes Z\, \otimes Z\,,\\
    &(e)& P_1\otimes P_1 \otimes P_1 \otimes \id \otimes \id\,\otimes
    \id\,,\\
    &(f)& \,\id\,\otimes\,\id\,\otimes\,\id\,\otimes P_2 \otimes P_2\otimes P_2,
\end{eqnarray*}
 where $P_1\!=\!|0\rangle\langle 0|\!+\!|1\rangle\langle
1|e^{2\pi i/3}$ and $P_2\!=\!e^{\pi i/2}|0\rangle\langle
0|\!+\!|1\rangle\langle 1|e^{\pi i}$. These local symmetries are
supplemented by the non-local joint permutation symmetry
\begin{displaymath}
    (g) \;\; {\cal P}(123)\times {\cal
    P}(456)\hspace*{1.5cm}
\end{displaymath}
where ${\cal P}$ represents an arbitrary index permutation. The
symmetries (a) - (g)  allow for a considerable simplification of
$\Omega$ and $\omega$. Indeed,  the symmetries (b), (c) and (e)
ensure that the matrix elements
$\Omega_{i_1j_1k_1l_1m_1n_1,i_2j_2k_2l_2m_2n_2}$ can only be
non-zero if the indices satisfy simultaneously $i_1\!=\!i_2$,
$j_1\!=\!j_2$ and $k_1\!=\!k_2$ or $i_1\!=\!1\!-\!i_2$,
$j_1\!=\!1\!-\!j_2$ and $k_1\!=\!1\!-\!k_2$. The symmetry (g)
yields
\begin{equation}
    \Omega_{abcdef,ghijkl} =
    \Omega_{{\cal P}(abc){\cal P}(def),{\cal P}(ghi){\cal P}(jkl)}
\label{perm}
\end{equation}
for any index permutation ${\cal P}$. Symmetry (a) yields
\begin{eqnarray}
    \Omega_{000l_1m_1n_1,000l_2m_2n_2} &=&
    \Omega_{111l_1m_1n_1,111l_2m_2n_2},\label{inv1}\\
    \Omega_{001l_1m_1n_1,001l_2m_2n_2} &=&
    \Omega_{110l_1m_1n_1,110l_2m_2n_2}, \label{inv2}\\
    \Omega_{000l_1m_1n_1,111l_2m_2n_2} &=&
    \Omega_{111l_1m_1n_1,000l_2m_2n_2}.\label{inv3}
\end{eqnarray}

 Presenting all nonzero matrix elements of
$\Omega_{abcdef,ghijkl}$ for $(abc,ghi)\!=\!(000,000)$,
$(abc,ghi)\!=\!(001,001)$ and $(abc,ghi)\!=\!(000,111)$ fixes all
other matrix elements by virtue of the symmetries  Eqs.\
(\ref{perm}-\ref{inv3}) and the Hermiticity of $\Omega$.  To
obtain a trial solution we chose
\begin{eqnarray*}
    \Omega_{000000,000000} &\!=\!& \Omega_{001000,001000} = -\Omega_{000000,111000} = b_1,\\
    \Omega_{000001,000001} &\!=\!& \Omega_{000001,000010} =
    \Omega_{000001,000100} = b_2,\\
    \Omega_{000011,000011} &\!=\!& \Omega_{000011,000101} =
    \Omega_{000011,000110} = b_4,\\
    \Omega_{001001,001001} &\!=\!& -\Omega_{001001,001010} =
    -\Omega_{001001,001100} = b_2,\\
    \Omega_{001010,001010} &\!=\!& \Omega_{001010,001100} =
    \Omega_{001100,001100} = b_2,\\
    \Omega_{001011,001011} &\!=\!& \Omega_{001011,001101} =
    -\Omega_{001011,001110} = -b_4,\\
    \Omega_{001101,001101} &\!=\!& -\Omega_{001101,001110} =
    \Omega_{001110,001110} = b_4,\\
    \Omega_{001111,001111} &\!=\!& 3\Omega_{000111,000111} = -3\Omega_{000111,111111} = 3b_6,\\
    \Omega_{000001,111001} &\!=\!& \Omega_{000001,111010} =
    \Omega_{000001,111100} = b_2,\\
    \Omega_{000010,111010} &\!=\!& \Omega_{000010,111100} =
    \Omega_{000100,111100} = b_2,\\
    \Omega_{000011,111011} &\!=\!& \Omega_{000011,111101} =
    \Omega_{000011,111110} = -b_4,\\
    \Omega_{000101,111101} &\!=\!& \Omega_{000101,111110} =
    \Omega_{000110,111110} = -b_4.
\end{eqnarray*}
Likewise,  the non-zero matrix elements of $\omega_V$ can be
constructed from
\begin{eqnarray*}
    (\omega_V)_{000,000} &=& 1-b_6-3b_4-3b_2-b_1, \\
    (\omega_V)_{001,001} &=& (\omega_V)_{000,111}, \\
    (\omega_V)_{000,111} &=& b_6+3b_4-3b_2+b_1,
\end{eqnarray*}
where we chose
\begin{eqnarray*}
    b_1 &=& \frac{1+\sqrt{1-4x^2}}{6},\;\;
    b_2 = \frac{x}{3},\;\;\; b_4 =\frac{b_2^2}{b_1},\;\;\; b_6 =
    \frac{9b_4^2}{3x},\\
    x &=& \frac{1}{8}(-2+(18-6\sqrt{3})^{1/3}+(18+6\sqrt{3})^{1/3}).
\end{eqnarray*}
A lengthy but elementary calculation  (preferably executed
employing a program capable of symbolic manipulations) then
confirms that this trial solution satisfies all the constraints
and yields the success probability
\begin{eqnarray}
    \hbox{tr}\{\Omega \rho_{GHZ} \otimes \rho_W\} = 6b_2.
\label{optsol}
\end{eqnarray}

We then consider the dual problem of the primal problem Eq.\ (\ref{optcom})
\cite{Boyd04a}. Every feasible point of the dual problem provides an upper
bound on the solution of the primal problem Eq.\ (\ref{optcom}). The above
result of Eq.\ (\ref{optsol}) is then proven to be optimal as shown in
Appendix \ref{sec: Proof of the optimality}.

As a consequence, the optimal probability for the transformation of a GHZ to
a $W$ state under PPT-operations is given by
\begin{equation}
    p(GHZ\rightarrow W) = 6b_2 \approx 0.75436...,
\label{optsol 2}
\end{equation}
that is more than 75\%. This very high success probability is
somewhat surprising, since the success probability for the LOCC
transformation is strictly zero.  Note that this result also
implies that a GHZ state can be transformed into a W state
employing LOCC supplemented by PPT-bound entanglement.
\section{Trace non-preserving CP-PPT maps}
\label{sec: Trace non-preserving}
 In the previous two sections we have demonstrated explicitly
that the success probability for the transformation between pure
states can in some cases be improved significantly by employing
PPT-operations instead of LOCC operations. Obtaining the optimal
success probabilities is a hard task, however, especially in the
multipartite setting. In the following we will consider the
slightly more tractable setting of trace non-preserving PPT maps.
In this setting we also optimize a CP-PPT map $\Psi$ or
equivalently the associated state $\Omega$, but the trace
preserving condition of Eq.\ (\ref{eq: trace-preserving}) is
replaced by the trace non-increasing condition of
\begin{equation}
    \hbox{tr}_{V'}\{\Omega\} \le \id_V.
\end{equation}
As a result, the completion  $\psi$  of the map $\Psi$ is a CP map
but  it  is {\em not} necessarily a PPT map.  This will generally
allow to find success probabilities for state transformations that
are larger than those obtained under trace-preserving PPT
operations. It is important to note, however, that any
transformation that possesses a non-vanishing success probability
under {\em trace non-preserving} CP-PPT maps will also have a
non-vanishing success probability under {\em trace preserving}
CP-PPT maps. To see this, let $\Omega(\Psi)$ be the state
corresponding to a trace non-preserving CP-PPT map $\Psi$. Since
the completion $\psi$ is not necessarily a PPT map,
$\omega(\psi)^{\Gamma_V}$ is sometimes a NPT-state. However, if we consider
the states of $\Omega'(\Psi')\!=\!\epsilon \Omega(\Psi)$ and
$\omega'(\psi')\!=\!\epsilon \omega(\psi)+
(1\!-\!\epsilon)\id\otimes\id/ {\dim\{{\cal H}(V')\}}$, the state
$(\omega')^{\Gamma_V}$ becomes a PPT-state for a nonzero value of
$1\!\ge\!\epsilon\!>\!0$.  Both $(\Omega')^{\Gamma_V}$ and
$(\omega')^{\Gamma_V}$ are PPT
states  satisfying the trace preserving condition of Eq.\
(\ref{eq: trace-preserving}), and $\Psi'$ accomplishes the same
transformation as $\Psi$  albeit with a smaller success
probability. In this way, one can always construct a trace
preserving CP-PPT map from the trace non-preserving CP-PPT map
giving the same transformation.

The optimal probability in the trace non-preserving scheme for the
transformation of maximally entangled states ($d'\!>\!d$)  can be
obtained in the same fashion as section III. Employing the
notation of section III we obtain the constraints
\begin{eqnarray*}
    1\ge a_1 \ge 0, \;\;\;\; a_2 \ge 0, \;\;\;\; a_4 \ge 0, \;\;\;\;
    1\ge a_2 + a_4, \hspace*{-0.75cm}&&  \\
    (d'+1)a_1+(d'+1)(d-1)a_2+(d-1) a_4&\ge& 0, \\
    -(d'+1)a_1+(d'+1)(d+1)a_2+(d+1)a_4&\ge& 0, \\
    -(d'-1)a_1-(d'-1)(d-1)a_2+(d-1)a_4&\ge& 0, \\
    (d'-1)a_1-(d'-1)(d+1)a_2+(d+1) a_4&\ge& 0
\end{eqnarray*}
under which the success probability, given by $a_1$, has to be
maximized. The result is
\begin{equation}
p(P^+_d \rightarrow P^+_{d'})= \frac{d-1}{d'-1},
\label{eq: probability of mes 2}
\end{equation}
whose PPT map $\Psi$ is, on the level of the state $\Omega$,
\begin{equation}
\Omega=\frac{d-1}{d'-1}P^+_d\otimes P^+_{d'}
+\frac{1}{d'^2-1}(\id-P^+_d)\otimes (\id-P^+_{d'}).
\label{eq: bipartite omega}
\end{equation}
 It is noteworthy that the probability of Eq.\ (\ref{eq:
probability of mes 2})  can be  written as a ratio of the
negativity of the  initial  and target state, i.e.
\begin{displaymath}
    p(P^+_d \!\rightarrow\! P^+_{d'}) =
    \frac{N(P^+_d)}{N(P^+_{d'})}
\end{displaymath}
where $N(\sigma)\!=\!(\hbox{tr}|\sigma^\Gamma|-1)/2$
\cite{Audenaert03a,Negativity}. This somewhat fascinating
expression resembles the case of the LOCC transformation of pure
states, where the optimal probability agrees with a ratio of a
LOCC-monotone such that the partial summation of squared Schmidt
coefficients \cite{Vidal99a}. Although the monotonicity of the
negativity in {\em trace non-preserving} PPT-operations has not
been proved yet (in {\em trace preserving} PPT operations with a
single outcome the negativity is a monotone \cite{Audenaert03a}),
the tractable expression of Eq.\ (\ref{eq: probability of mes 2})
is likely to be explained as a ratio of some monotone function.

In the tripartite setting, the optimization of the success probability is
still a hard task even in this trace non-preserving scheme.
The result of the optimization for the transformation of
$\hbox{GHZ}\!\rightarrow\!W$ is
\begin{equation}
    p(GHZ\rightarrow W) = \frac{4}{5},
\end{equation}
and for the transformation of $W\!\rightarrow\!\hbox{GHZ}$ we have
\begin{equation}
    p(W\rightarrow GHZ)=\frac{1}{3}.
    \label{eq: probability from W to GHZ}
\end{equation}
The proof for these two results are described in appendices
\ref{sec: GHZ to W non-trace preserving} and \ref{sec: W to GHZ
non-trace preserving}. This result implies that the transformation
of $W\!\rightarrow\!\hbox{GHZ}$ is also possible by trace
preserving PPT-operations, although the optimal probability may be
smaller than 1/3. Therefore, PPT-operations can inter-convert even
the LOCC-incomparable pure states. In the next section,  we will
completely clarify the convertibility by PPT-operations for all
multipartite pure states in the single copy setting.

\section{Convertibility of pure states}
\label{sec: Convertibility of pure states}
 In this section we will consider the transformation between
single copies of $N$-partite pure states under PPT-operations. By
definition, PPT-operations map PPT-states to PPT-states. As a
consequence, transformations such as
$|\phi^+_{AB}\rangle\!\otimes\!|0_C\rangle\!\rightarrow\!|GHZ\rangle$
or $|\phi^+_{AB}\rangle\!\otimes\!|0_C\rangle\!\rightarrow\!
|0_A\rangle\!\otimes\!|\phi^+_{BC}\rangle$ are impossible, since
they are not PPT-preserving with respect to party $C$.
Therefore, let us first assume for the transformation of
$|\psi\rangle\!\rightarrow\!|\phi\rangle$ that both $|\psi\rangle$
and $|\psi\rangle$ are ``genuinely'' entangled over all $N$
parties. This assumption means that
\begin{equation}
    (|\psi\rangle\langle\psi|)^{\Gamma_i}\not\ge0 \hbox{~~~and~~~}
    (|\phi\rangle\langle\phi|)^{\Gamma_i}\not\ge0,
    \label{eq: genuinely entangled}
\end{equation}
for all possible bipartite partitioning of $i$. For example,
$i\!=\!A$, $B$, $C$ in  a tripartite setting, and $i\!=\!A$, $B$,
$C$, $D$, $AB$, $AC$, $AD$ in  a four-partite setting. As
discussed in the previous section, it suffices to consider trace
non-preserving CP-PPT maps  $\Psi$  in order to check the
convertibility under  trace preserving  PPT-operations. Therefore,
we will construct  an  $\Omega$ satisfying the constraints
\begin{eqnarray}
    &&
    \hbox{tr}\{\Omega |\psi\rangle\langle\psi|
    \otimes(\id-|\phi\rangle\langle\phi|)\} = 0, \nonumber \\
    && \;\; \Omega^{\Gamma_{V}} \ge 0, \;\;\;
    (\Omega^{\Gamma_{V_i}\otimes\Gamma'_{V_i}})^{\Gamma_{V}} \ge 0,
    \label{eq: Multipartite constraints}
\end{eqnarray}
where $|\psi\rangle\!\in\!{\cal H}(V)$, $|\phi\rangle\!\in\!{\cal
H}(V')$, and $i$  stands for any  possible bipartite partitioning
as explained below Eq. (\ref{eq: genuinely entangled}). We have
omitted the trace non-increasing condition, because we are not
interested in the explicit value of the success probability but
only whether it is zero or not. In view of Eq.\ (\ref{eq:
bipartite omega}),  a suitable  trial form is
\begin{equation}
    \Omega=x|\psi\rangle\langle\psi|\otimes |\phi\rangle\langle\phi|
    +(\id-|\psi\rangle\langle\psi|)\otimes(\id-|\phi\rangle\langle\phi|),
    \label{eq: Trial omega}
\end{equation}
for which the first two constraints in Eq.\ (\ref{eq: Multipartite
constraints}) are satisfied when $x\!\ge\!0$. Furthermore, due to
the assumption of Eq.\ (\ref{eq: genuinely entangled}), the last
constraint
$(\Omega^{\Gamma_i\otimes\Gamma'_i})^{\Gamma_V}\!\ge\!0$ is also
satisfied for an appropriate value of $x\!=\!x_0\!>\!0$ as shown
in \cite{Ishizaka04b}. As a result, for $x\!=\!x_0$ we have
\begin{equation}
    \hbox{tr}\{\Omega(|\psi\rangle\langle\psi|\otimes|\phi\rangle\langle\phi|)\}
    =x_0>0,
\end{equation}
 so that for arbitrary pairs of genuine $N$-partite entangled
states of $|\psi\rangle$ and $|\phi\rangle$ we can always find an
$\Omega$ such that
$p(|\psi\rangle\!\rightarrow\!|\phi\rangle)\!>\!0$.  As a
consequence, all genuine $N$-partite pure entangled states are
inter-convertible by PPT-operations.
In this way, the classification of $N$-partite entanglement is
drastically simplified when we consider PPT-operations.

\begin{figure}
\centerline{\scalebox{0.45}[0.45]{\includegraphics{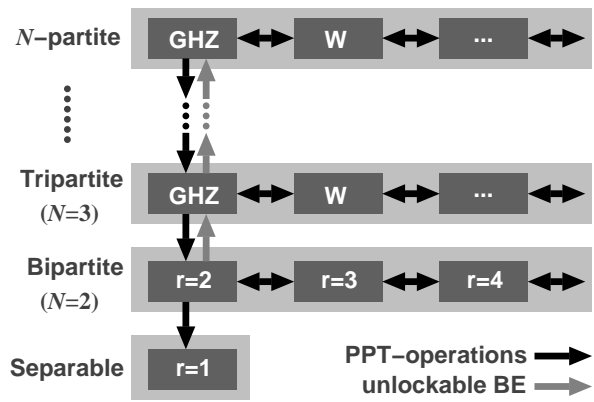}}}
\caption{
The classification and convertibility of multipartite pure entangled states
under PPT-operations.
$r$ denotes the Schmidt rank of bipartite entanglement, and
the set of entangled parties in $(N\!-\!1)$-partite entanglement is assumed to
be a subset of the set of entangled parties in $N$-partite entanglement.
There is only one type of $N$-partite entanglement under PPT-operations.
Furthermore, the convertibility with the support of unlockable bound
entanglement (BE) is also shown (see also \protect{\cite{Dur99a}}).
}
\label{fig: multipartite}
\end{figure}

Let us next investigate the convertibility between a $N$-partite
state $|\psi^{(N)}\rangle$ and a $(N\!-\!1)$-partite state
$|\phi^{(N-1)}\rangle$. It is obvious that
$|\phi^{(N-1)}\rangle\!\rightarrow\!|\psi^{(N)}\rangle$ is
impossible because  such a transformation is not PPT-preserving.
Likewise the transformation of
$|\psi^{(N)}\rangle\!\rightarrow\!|\phi^{(N-1)}\rangle$ is
impossible if the  set of entangled parties  in
$|\phi^{(N-1)}\rangle$  is not a subset of the  set of entangled
parties in $|\psi^{(N)}\rangle$ (e.g.
$|\psi^{(3)}_{ABC}\rangle\!\rightarrow\!|\phi^{(2)}_{AD}\rangle$
is impossible). Otherwise, the transformation is possible because
a $N$-partite GHZ state can be transformed to a
$(N\!-\!1)$-partite GHZ state by LOCC, and hence the sequential
transformation of
$|\psi^{(N)}\rangle\!\rightarrow\!|GHZ^{(N)}\rangle\!\rightarrow\!
\!|GHZ^{(N-1)}\rangle\!\rightarrow\!|\phi^{(N-1)}\rangle$ is
possible. The classification and convertibility of arbitrary
multipartite pure entangled states under PPT-operations are
summarized in Fig.\ \ref{fig: multipartite}.

It is important to note here that the power of PPT-operations, by
which $N$-partite pure entangled states becomes inter-convertible
as discussed above immediately implies that the same holds for LOCC
supported by PPT bound entanglement. This is due to the fact that any
PPT-transformation can be accomplished (with smaller but nonzero probability)
by LOCC supported by the  additional resource of PPT-states
\cite{Cirac01a} (see the note of \cite{Note_for_PPT_operations}). Indeed,
\begin{equation}
    \Psi(\rho)=\hbox{tr}_V \{ \Omega(\Psi)^{\Gamma_V} (\rho^{\Gamma_V} \otimes \id) \},
\end{equation}
and the state of $\Omega(\Psi)^{\Gamma_V}\!\ge\!0$, which is a
PPT-state if $\Psi$ is a CP-PPT map due to
$(\Omega(\Psi)^{\Gamma_i\otimes\Gamma'_i})^{\Gamma_V}\!\ge\!0$, is
utilized and consumed in the LOCC-implementation of $\Psi(\rho)$
\cite{Cirac01a}.
If a CP-PPT map $\Psi$ can accomplish  a
transformation that is impossible under LOCC alone, then
$\Omega(\Psi)^{\Gamma_V}$ must be entangled (otherwise the
transformation  can also be accomplished by LOCC because LOCC
 can generate any separable state), and  therefore
the PPT-state $\Omega(\Psi)^{\Gamma_V}$ is a PPT bound entangled
state \cite{Horodecki98a}. Consequently, one can conclude that the
transformation such as $\hbox{GHZ}\!\leftrightarrow\!W$ can be
accomplished by LOCC with the consumption of PPT bound entangled
states. Much attention has been paid to bound entanglement to
clarify its properties, and several  applications  of bound
entanglement have been reported
\cite{Horodecki99c,Dur99a,Smolin01a,Shor03a,Murao01a,Dur01a,Kaszlikowski02a,
Sen02a,Dur01b,Horodecki03a,Dur04a,Augusiak04a}.
As shown above, PPT bound entanglement enables the LOCC implementation
of large classes of entanglement transformations that are impossible by
LOCC alone.
\section{Unlockable states and conversion of pure states}
\label{sec: Unlockable states}
As mentioned in the previous section, the transformation
\begin{equation}
    |\phi^+_{AB}\rangle\otimes|0_C\rangle \rightarrow |GHZ_{ABC}\rangle
    \label{eq: EPR to GHZ}
\end{equation}
cannot be achieved even when PPT-operations are employed,
and therefore cannot be achieved by LOCC supported by PPT-bound entanglement.
However, it has been shown that a GHZ state can be distilled from
a tripartite NPT-bound entangled state, if $A$ and $B$ perform a global
operation on the state \cite{Dur99a}.
Such NPT-bound entangled states are called unlockable states
because bound entanglement is unlocked by the global operation
\cite{Smolin01a,Wei04a,Augusiak04b}.
The global operation of $A$ and $B$ can be accomplished by LOCC
consuming $|\phi^+_{AB}\rangle$, and consequently the transformation
of Eq.\ (\ref{eq: EPR to GHZ}) is possible when LOCC is supported by
the unlockable bound entanglement \cite{Dur99a}.
Likewise, unlockable states which can be utilized for the
LOCC-transformation from a $(N\!-\!1)$-partite GHZ state
to a $N$-partite GHZ state have been shown in \cite{Dur99a}.
In this section we consider this type of transformation using a certain
general scheme.

To this end, we generalize PPT-operations by relaxing the
PPT-preserving condition with respect to $C$,
$(\Omega^{\Gamma_{V_C}\otimes\Gamma'_{V_C}})^{\Gamma_{V}}\!\ge\!0$,
which is responsible for the impossibility of the transformation
of Eq.\ (\ref{eq: EPR to GHZ}). We will therefore construct a map
$\Psi$ whose associated Hermitian operator $\Omega$ satisfies
\begin{eqnarray}
&&
\hbox{tr}\{\Omega P^+_{AB} \otimes(\id-\rho_{GHZ})\} = 0, \;\;
\Omega^{\Gamma_{V}} \ge 0, \nonumber \\
&&
(\Omega^{\Gamma_{V_A}\otimes\Gamma'_{V_A}})^{\Gamma_{V}} \ge 0, \;\;
(\Omega^{\Gamma_{V_B}\otimes\Gamma'_{V_B}})^{\Gamma_{V}} \ge 0,
\label{eq: Multipartite constraints 2}
\end{eqnarray}
where
$P^+_{AB}\!=\!|\phi^+_{AB}\rangle\langle\phi^+_{AB}|\!\in\!{\cal
H}(V)$ and $\rho_{GHZ}\!\in\!{\cal H}(V')$.
As a trial form for $\Omega$, we adopt again Eq.\ (\ref{eq: Trial omega}), i.e.
\begin{equation}
\Omega=xP^+_{AB} \otimes \rho_{GHZ}
+(\id-P^+_{AB})\otimes(\id-\rho_{GHZ}).
\label{eq: Unlockable state}
\end{equation}
As mentioned in the previous section,
\[
(P^+_{AB})^{\Gamma_A}\not\ge0 \hbox{~~~and~~~} \rho_{GHZ}^{\Gamma_A}\not\ge0
\]
ensure the existence of $x_0\!>\!0$ such that
$(\Omega^{\Gamma_{V_A}\otimes\Gamma'_{V_A}})^{\Gamma_{V}}\!\ge\!0$
for $0\!<\!x\!\le\!x_0$ (indeed, we have $x_0\!=\!3$), and
likewise with respect to $B$. We can now check easily that all
constraints in Eq.\ (\ref{eq: Multipartite constraints 2}) are
satisfied for $x\!=\!3$,  yielding a nonzero success probability
$p(|\phi^+_{AB}\rangle\!\rightarrow\!|GHZ\rangle)\!>\!0$ since
$\hbox{tr}\{\Omega(P^+_{AB}\otimes\rho_{GHZ})\}\!>\!0$.
Consequently, the transformation of
$|\phi^+_{AB}\rangle\!\rightarrow\!|GHZ\rangle$ is possible under
the set of operations that maps NPT-BE states with
respect to party $C$ into itself, as expected. Employing
symmetries of $P^+_{AB}\otimes\rho_{GHZ}$, the optimized success
probability in the trace non-preserving scheme is then obtained as
\begin{equation}
p(|\phi^+_{AB}\rangle\!\rightarrow\!|GHZ\rangle)=\frac{3}{5},
\end{equation}
and, on the level of states, the map $\Psi$ realizing
this success probability is given by
\begin{eqnarray}
\Omega&=&\frac{3}{5}P^+_{AB} \otimes \rho_{GHZ} \nonumber\\
&+&\frac{1}{5}(\id-P^+_{AB})\otimes(\id-\rho_{GHZ}
\!-\!\rho_{001}-\rho_{110}),
\label{eq: Optimized unlockable state}
\end{eqnarray}
where $\rho_{001}\!=\!|001\rangle\langle001|$ and
$\rho_{110}\!=\!|110\rangle\langle110|$.
It can be confirmed that the state $\Omega^{\Gamma_V}$ is an unlockable state
as follows.
Due to the constraints of
$(\Omega^{\Gamma_{V_A}\otimes\Gamma'_{V_A}})^{\Gamma_{V}}\!\ge\!0$
and
$(\Omega^{\Gamma_{V_B}\otimes\Gamma'_{V_B}})^{\Gamma_{V}}\!\ge\!0$,
the mixed state of $\Omega^{\Gamma_V}$ is undistillable by LOCC,
because LOCC is PPT-preserving and  no tripartite and bipartite
{\em pure} entangled state exists that is PPT with respect to
both $A$ and $B$. However, a GHZ state can be distilled from $\Omega^{\Gamma_V}$ of
Eq.\ (\ref{eq: Unlockable state}) or Eq.\ (\ref{eq: Optimized unlockable state}),
if $A$ and $B$ perform global operations that
distinguish $P^+_{AB}$ and $\id\!-\!P^+_{AB}$.

Similarly, the map $\Psi$, whose
associated state is
\begin{equation}
    \Omega=3\rho_{GHZ}^{(N)}\otimes \rho_{GHZ}^{(N')}
    +(\id\!-\!\rho_{GHZ}^{(N)})\otimes (\id\!-\!\rho_{GHZ}^{(N')}),
\label{eq: Multipartite unlockable state}
\end{equation}
can transform a $N$-partite GHZ state ($\rho_{GHZ}^{(N)}$) to a
$N'$-partite GHZ state, and furthermore the state
$\Omega^{\Gamma_V}$ is an unlockable state if $N'\!>\!N\!\ge\!2$
\cite{Note_for_unlockable_state}. As shown in the previous
section, all genuine $N$-partite entangled states are
inter-converted by PPT maps. The composition of the PPT maps and
the map given in 
Eq.\ (\ref{eq: Multipartite unlockable state}) is again a map 
whose associated state is an unlockable state.
This implies that all pure entangled states can be inter-converted
independently of the number of parties ($N$) when a single copy of
an appropriate unlockable bound entangled state is available as a
resource. In this way, the consumption of unlockable bound
entanglement allows to overcome the LOCC-constraint between pure
states with different sets of entangled  parties, while the
consumption of PPT bound entanglement overcomes the
LOCC-constraint between pure states with the same set of entangled
parties (Fig.\ \ref{fig: multipartite}).
\section{Single copy distillation}
\label{sec: Single copy distillation}
 So far, we have concentrated our attention on the discussion
of transformations between  pure states. In this section, we
 will now consider  the transformation of a single copy of
a mixed state  $\rho$ into a maximally entangled state $P^+_{d'}$,
i.e. the single copy distillation from a mixed state
 employing PPT-operations.

 Let us consider the antisymmetric Werner state which is
defined as
\begin{equation}
    \sigma^a_d = \frac{2}{d^2-d}P^a_d
    =\frac{2}{d^2-d}\sum_{j>i} |\psi^-_{ij}\rangle\langle\psi^-_{ij}|,
\label{eq: Antisymmetric Werner state}
\end{equation}
 where $P^a_d$ is the projector onto the antisymmetric
subspace of ${\mathbb C}^d\otimes{\mathbb C}^d$, and
$|\psi^-_{ij}\rangle\!=\!(|ij\rangle\!-\!|ji\rangle)/\sqrt{2}$.
For the transformation of $\sigma^s_d\!\rightarrow\!P^+_{d'}$, we
can construct CP-PPT maps of $\Psi$ and its  CP-PPT completion
$\psi$ employing the twirling symmetries of the two states.
The result of the optimization is, on the level of the state $\Omega$
(the state $\omega$ is given by $\omega_V\!=\!\id\!-\!\hbox{tr}_{V'}\Omega$),
\[
    \Omega= \frac{2}{dd'+d'-2d} \Big[P^a_d \otimes P^+_{d'}
    + (d'-1) P^s_d \otimes \frac{\id-P^+_{d'}}{{d'}^2-1}\Big]
\]
for $d'\!\ge\!d\!\ge\!2$, and
\begin{eqnarray*}
    \Omega&=& \frac{2}{d(d'-1)} \Big[ P^a_d
           +  \frac{(d-d')}{(d+1)d'} P^s_d \Big] \otimes P^+_{d'} \\
    &+& \frac{2(d'+1)}{(d+1)d'} P^s_d \otimes \frac{\id-P^+_{d'}}{{d'}^2-1}
\end{eqnarray*}
for $2\!\le\!d'\!\le\!d$ where $P^s_d$ is the projector onto the symmetric
subspace of ${\mathbb C}^d\otimes{\mathbb C}^d$.
The optimal success probability  under trace preserving
CP-PPT-operations is then given by
\begin{equation}
    p(\sigma^a_d\rightarrow P^+_{d'})=
    \left\{\begin{array}{lc}
    \displaystyle\frac{2}{dd'+d'-2d} & \hbox{~for $d'\!>\!d\!\ge\!2$,}
    \\[0.3cm]
    \displaystyle\frac{2}{d(d'-1)}     & \hbox{~for $2\!\le\!d'\!\le\!d$.}
    \end{array}\right.
    \label{eq: probability of aws}
\end{equation}
Therefore, the success probability is nonzero for
$d'\!\ge\!2$.

On the other hand, the success probability for the  same
transformation under LOCC operations alone is strictly zero
whenever  $d'\!>\!2$. This  can be proven  as follows:
 The $|\psi^-_{ij}\rangle$ in Eq.\ (\ref{eq: Antisymmetric
Werner state}) are maximally entangled states on ${\mathbb
C}^2\otimes{\mathbb C}^2$.  Therefore, each $|\psi^-_{ij}\rangle$
can be prepared from $P^+_2$ by local unitary  transformations
only.  As $\sigma^a_d$ is an equal mixture of  all possible
$|\psi^-_{ij}\rangle$, $\sigma^a_d$ can be prepared from a single
copy of $P^+_2$ by LOCC, and hence the transformation of
$P^+_2\!\rightarrow\!\sigma^a_d$  has a finite success
probability.  If we furthermore assume that for $d'\!>\!2$ the
transformation $\sigma^a_d\!\rightarrow\!P^+_{d'}$ has a finite
success probability under LOCC, then this implies that
$P^+_2\!\rightarrow\!\sigma^a_d\!\rightarrow\!P^+_{d'}$ also has a
finite success probability under LOCC. This contradicts that the
Schmidt rank cannot be increased by LOCC. Therefore, the result of
Eq.\ (\ref{eq: probability of aws}) implies that the success
probability of the single copy distillation is also significantly
improved when PPT-operations are considered.

It should be noted  that the transformation of
$\sigma^a_d\!\rightarrow\!P^+_2$ is possible  under LOCC. Indeed,
the local projection $P\!\otimes\!P$ to $\sigma^a_d$, where
$P\!=\!|0\rangle\langle0|\!+\!|1\rangle\langle1|$, can accomplish
this. Furthermore, $P^+_2\!\rightarrow\!P^+_{d'}$ is possible
under PPT-operations, which enables the sequential transformation
of $\sigma^a_d\!\rightarrow\!P^+_2\!\rightarrow\!P^+_{d'}$.
Therefore, the  feasibility of
$p(\sigma^a_d\!\rightarrow\!P^+_{d'})$ can be regarded as being a
consequence of the  feasibility of
$p(P^+_2\!\rightarrow\!P^+_{d'})$  under PPT-operations.  Note
however, that Eqs.\ (\ref{eq: probability of mes}) and (\ref{eq:
probability of aws}) for $d'\!>\!2$ imply that we have
\begin{equation}
    p(\sigma^a_d\!\rightarrow\!P^+_{d'})>p(\sigma^a_d\!\rightarrow\!P^+_2)
    p(P^+_2\!\rightarrow\!P^+_{d'}).
\end{equation}
 Hence the direct transformation is accomplished with a higher
success probability than that for the corresponding sequential
transformation.

 The discussion above demonstrates that PPT-operations can
improve the success probability of the single copy distillation
for some mixed states. One may  perhaps expect that  single copy
distillation becomes possible for all NPT mixed states when we
consider PPT-operations.  This, however, is not the case. As shown
in \cite{Kent98a} (see also \cite{Horodecki99a}),
LOCC cannot distill any pure entangled state
from a single copy of mixed states $\rho$ on ${\mathbb
C}^d\otimes{\mathbb C}^d$ if $\hbox{rank}(\rho)\!\ge\!d^2\!-\!2$.
For such  high rank  mixed states, PPT-operations  cannot distill
any pure entangled state  either. The proof  of this statement
 is given in Appendix \ref{sec: Single copy distillation from
high rank mixed states}.

 This highlights the fact that LOCC state manipulation suffers
certain restrictions that PPT-operations cannot relax. Indeed, the
convertibility of some mixed states (into pure entangled states)
at the single copy level, and therefore the convertibility of
mixed states under PPT-operations  remains much more involved than
the convertibility of pure states.
\section{Summary}
\label{sec: Summary}
 In this paper we have considered the transformation of single
copies of multi-particle entanglement under sets of operations that
are larger than the class of local operations and classical
communication (LOCC). In particular, we considered probabilistic
state transformations under positive partial transpose preserving
maps (PPT-maps). We demonstrated that transformations that are
strictly impossible under LOCC can have a finite success
probability under trace preserving PPT-maps. For specific examples
the optimal success probabilities are determined. Surprisingly
large values are obtained for example for the transformation from
the GHZ to $W$ state which under trace preserving PPT-maps has a
success probability of more than 75\% while it is strictly
forbidden under LOCC. Furthermore, we completely clarified the
convertibility of arbitrary multipartite pure states under
PPT-operations. As a remarkable result, we showed that all
$N$-partite pure entangled states are inter-convertible under
PPT-operations at the single copy level, and therefore infinitely
many different types of entanglement under LOCC are merged into
only one type. In this way, a drastic simplification in the
classification of pure state entanglement occurs when the
constrained set of operations is changed from LOCC to
PPT-operations. It should be emphasized that despite such drastic
simplification in the single copy settings, the theory of
entanglement under PPT-operations possesses the desirable
properties that PPT-operations alone cannot create pure state
entanglement and that the amount of bipartite pure state
entanglement is uniquely determined in asymptotic settings
\cite{Note_for_entanglement_measure}.

The above results can be regarded as an application of PPT-bound
entanglement. In multipartite settings however another type of
bound entanglement called unlockable bound entanglement exists.
Motivated by this, we enlarged the class of PPT-operations to
consider the effects of unlockable bound entanglement. As a result
we showed that all pure entangled states become inter-convertible
independent of the number of parties, and therefore a further
drastic simplification in the classification of pure states occurs
when LOCC is supported by unlockable bound entanglement.

Finally, we considered  one aspect of mixed state entanglement
transformations, namely  the single copy distillation by
PPT-operations. We  demonstrated that PPT-operations can distill a
pure entangled state from a single copy of some mixed states with
finite success probability, while the success probability under
LOCC is strictly zero. However, we also proved that PPT-operations
cannot distill  pure entangled state from mixed states with very
high rank. Therefore, certain restrictions of entanglement
manipulation of mixed states under LOCC persist under PPT-maps,
and the classification of mixed states under PPT-operations in the
single copy settings is not as simple as that in the pure state
case.

It is important to further clarify how the structure of theory of
entanglement is simplified under PPT-operations especially in the
mixed state settings and in asymptotic settings,  as this might
enable a unified and systematic understanding of characteristics
of quantum entanglement as a resource.

\begin{acknowledgments}
This research was initiated during two visits to the ERATO project
on {\it Quantum Information Science}. This work is part of the
QIP-IRC (www.qipirc.org) supported by EPSRC (GR/S82176/0) and the
EU (IST-2001-38877), a Royal Society Leverhulme Trust Senior
Research Fellowship and the Leverhulme Trust.
\end{acknowledgments}

\appendix
\section{Optimality of the conversion from GHZ to $W$ state}
\label{sec: Proof of the optimality}
In this appendix, we prove the optimality of Eq.\ (\ref{optsol}),
the probability for the transformation from GHZ to $W$ state. To
this end, we consider the dual problem of the primal problem Eq.\
(\ref{optcom}) \cite{Boyd04a}.  The Lagrange function for the
minimization problem in Eq. (\ref{optcom}) is given by
\begin{eqnarray}
    L &=& -\hbox{tr}\{\Omega\rho_{GHZ}\otimes\id\}- \sum_{i=A,B,C} (\hbox{tr}\{\lambda_i^{\Gamma_i}\Omega\}
    +\hbox{tr}\{\mu_i^{\Gamma_i}\omega\})\nonumber\\
    &&\hspace*{-0.75cm}+ \hbox{tr}\{\lambda_{e} (\hbox{tr}_{V'}(\Omega+\omega) - \id)\}
    + \nu \hbox{tr}\{\Omega \rho_{GHZ}\otimes(\id-\rho_{W})\}\nonumber\\
    &&\hspace*{-0.75cm}+ \hbox{tr}\{\lambda_p (\hbox{tr}_{V'}\Omega - \id)\}
    + \hbox{tr}\{\lambda_{ep} (\hbox{tr}_{V'}\omega - \id)\}, \nonumber
\end{eqnarray}
where
$\lambda_p,\lambda_{ep},\lambda_A,\lambda_B,\lambda_C,\mu_A,\mu_B,\mu_C\ge
0$. This Lagrange function has to be minimized over all
$\Omega,\omega\ge 0$. This is feasible only if
\begin{eqnarray*}
    &&\!\!\!\!\!\!0\ge\!\!\!\!\! \sum_{i=A,B,C}\!\!\!\!\!\lambda_i^{\Gamma_i}\!
    +\!(\rho_{GHZ}\!-\!\lambda_p\!-\!\lambda_e)\otimes\id
    \!-\!\nu\rho_{GHZ}\otimes(\id\!-\!\rho_{W}), \nonumber\\
    &&\!\!\!\!\!\!0\ge\!-(\lambda_{e}+\lambda_{ep})\otimes\id + \mu_A^{\Gamma_A}
    +\mu_B^{\Gamma_B}+\mu_C^{\Gamma_C},
\end{eqnarray*}
in which case we obtain the dual function
\begin{equation}
    g(\lambda_p,\lambda_{ep},\lambda_e,\nu)
    = -\hbox{tr}\{\lambda_{ep}+\lambda_e+\lambda_p\}.
\end{equation}
 Every feasible point of the dual problem provides an upper
bound on the solution of the primal problem Eq.\ (\ref{optcom}).
With the symmetries shown in Sec.\ \ref{sec: Tripartite}, the
Lagrange dual problem of primal problem Eq.\ (\ref{optcom}) is
\begin{equation}
    \min \,\hbox{tr}\{\lambda_{ep}+\lambda_e+\lambda_p\}
\label{dual}
\end{equation}
under the constraints
\begin{eqnarray*}
(\rho_{GHZ}\!-\!\lambda_p\!-\!\lambda_e)\!\otimes\!\id
    \!-\!\nu\rho_{GHZ}\!\otimes\!(\id\!-\!\rho_{W})\!+ \!\!\!\!
    \sum_{i=A,B,C}\!\!\!\!\lambda_i^{\Gamma_{V_i}}\!\le\!0, \\
\lambda_A,\lambda_B,\lambda_C,\mu_A,\mu_B,\mu_C,\lambda_p,\lambda_{ep}
\ge0, && \qquad \\
 -(\lambda_{e}+\lambda_{ep})\otimes\id + \mu_A^{\Gamma_{V_A}}
    +\mu_B^{\Gamma_{V_B}}+\mu_C^{\Gamma_{V_C}} \le 0. && \qquad
\end{eqnarray*}
To prove the optimality of Eq.\ (\ref{optsol}), it suffices to provide a trial
solution for the dual problem that matches the value Eq.\ (\ref{optsol}).
To this end, we chose $\nu\!=\!\frac{8}{3}$,
$\lambda_{pe}\!=\!\lambda_p\!=\!0$, and
$(\lambda_{e})_{i,j}\!=\!0$ except for
\[
    (\lambda_{e})_{i,i} = b_2, \;\;\;
    (\lambda_{e})_{1,8} = (\lambda_{e})_{8,1} = -3b_2.
\]
Furthermore,
\begin{eqnarray*}
 (\mu_{A})_{i,i} &=& (\mu_{A})_{i+40,i+40} =
  -(\mu_{A})_{i,i+40} = \\
-(\mu_{A})_{i+40,i} &=& (\mu_{B})_{i+8,i+8} =
  (\mu_{B})_{i+32,i+32} = \\
-(\mu_{B})_{i+8,i+32} &=& -(\mu_{B})_{i+32,i+8} =
  (\mu_{C})_{i+16,i+16} = \\
(\mu_{C})_{i+24,i+24} &=& -(\mu_{C})_{i+16,i+24} =
  -(\mu_{C})_{i+24,i+16} = b_2
\end{eqnarray*}
for $i=9,\ldots,16$. Finally, one chooses the matrices
$\lambda_A^{\Gamma_{V_A}}$, $\lambda_B^{\Gamma_{V_B}}$ and
$\lambda_C^{\Gamma_{V_C}}$. As $\lambda_B^{\Gamma_{V_B}}$ and
$\lambda_C^{\Gamma_{V_C}}$ can be obtained from
$\lambda_A^{\Gamma_{V_A}}$ by cyclic permutations, we only need to
specify $\lambda_A^{\Gamma_{V_A}}$. For $i,j\!=\!1,\ldots,8$ we
have
\begin{eqnarray*}
    &&(\lambda_A^{\Gamma_{V_A}})_{i,j}=(\lambda_A^{\Gamma_{V_A}})_{56+i,56+j}=X_{i,j}, \\
    && (\lambda_A^{\Gamma_{V_A}})_{i,56+j}= (\lambda_A^{\Gamma_{V_A}})_{56+i,j}= Y_{i,j}, \\
    && (\lambda_A^{\Gamma_{V_A}})_{i+8,j+8}= (\lambda_A^{\Gamma_{V_A}})_{48+i,48+j}= \delta_{i,j},
\end{eqnarray*}
 where the nonzero elements of $X$ and $Y$ are given by
\begin{eqnarray*}
    X_{1,1} &=& 1, \;\; X_{4,4} = X_{6,6} = X_{6,4} = X_{4,6} = 25/16, \\
    X_{2,3} &=& X_{2,5} = X_{3,2} =  X_{5,2} = -5/4, \\
    Y_{1,1} &=& Y_{4,4} = Y_{6,6} = -1/3, \\
    Y_{2,2} &=& - Y_{7,7} = -1, \\
    Y_{2,3} &=& Y_{2,5} = Y_{3,2} = Y_{3,3} = Y_{3,5} = -Y_{4,6} = -2/3, \\
    Y_{5,2} &=& Y_{5,3} = -Y_{5,5} = -Y_{6,4} = -Y_{8,8} = -2/3, \\
    Y_{6,7} &=& Y_{7,6} = 7/80, \\
    Y_{7,4} &=& Y_{4,7} = (-42+\sqrt{159559})/1200.
\end{eqnarray*}
A direct calculation, ideally employing a software capable of
symbolic manipulations, now shows that these values determine a
feasible point of the dual problem. The dual function for the
above  choice yields the value $6b_2$, i.e. the same as for the
primal problem which establishes the optimality of the solution
for the primal problem.

\section{From GHZ to $W$ employing non-trace preserving PPT maps}
\label{sec: GHZ to W non-trace preserving}
In this appendix we determine the optimal success probability for
the transformation of a GHZ state to a W state under non-trace
preserving CP-PPT maps. This problem is equivalent to the
maximization of
\begin{eqnarray}
    \hbox{tr}\{\Psi(\rho_{GHZ})\} &=& \hbox{tr}\{\Omega
    \rho_{GHZ}\otimes\id\} \label{eq6}
\end{eqnarray}
under the constraints
\begin{eqnarray}
    \hbox{tr}\{\Omega\rho_{GHZ}\otimes(\id-\rho_{W})\} &\!=\!& 0, \nonumber\\
    \Omega^{\Gamma_{V}} \ge 0, \; \hbox{tr}_{V'}\{\Omega(\Psi)\} &\!\le\!& \id,
    \label{constraintsGHZW}\\
    (\Omega^{\Gamma_{A}\otimes\Gamma'_{A}})^{\Gamma_{V}}
    \ge  0,\;
    (\Omega^{\Gamma_{B}\otimes\Gamma'_{B}})^{\Gamma_{V}}
    &\!\ge\!& 0, \;
    (\Omega^{\Gamma_{C}\otimes\Gamma'_{C}})^{\Gamma_{V}}
    \ge  0.\nonumber
\end{eqnarray}
This problem possesses the same symmetries (a) - (g) presented in
section \ref{sec: Tripartite}. Following the same arguments as in
section \ref{sec: Tripartite} most matrix elements of $\Omega$
vanish. In the following we will present those non-vanishing
matrix elements that are sufficient to reconstruct all the
remaining non-zero elements of the trial solution from the
symmetries of the problem. With $b_1=0.8/3=2b_2=4b_3$ we find
\begin{eqnarray*}
    \Omega_{001000,001000} &\!=\!& 2\Omega_{001111,001111} = b_1,\\
    \Omega_{001001,001001} &\!=\!& \Omega_{001010,001010} = \Omega_{001100,001100} = b_2,\\
    \Omega_{001001,001010} &\!=\!& \Omega_{001001,001100} = -\Omega_{001010,001100} = -b_2,\\
    \Omega_{001011,001011} &\!=\!& \Omega_{001101,001101} = \Omega_{001110,001110} = b_3,\\
    \Omega_{001011,001101} &\!=\!& -\Omega_{001011,001110} = -\Omega_{001101,001110} =
    b_3,\\
    \Omega_{000000,000000} &\!=\!& 8\Omega_{000111,000111} = b_1,\\
    \Omega_{000001,000001} &\!=\!& \Omega_{000010,000010} = \Omega_{000100,000100} = b_2,\\
    \Omega_{000001,000010} &\!=\!& \Omega_{000001,000100} = \Omega_{000010,000100} = b_2,\\
    \Omega_{000011,000011} &\!=\!& \Omega_{000101,000101} = \Omega_{000110,000110} = b_3,\\
    \Omega_{000011,000101} &\!=\!&  \Omega_{000011,000110} = \Omega_{000101,000110} =    b_3,\\
    \Omega_{000000,111000} &\!=\!& 8\Omega_{000111,111111} = -b_1,\\
    \Omega_{000001,111001} &\!=\!& \Omega_{000010,111010} = \Omega_{000100,111100} = b_2,
\end{eqnarray*}
\begin{eqnarray*}
    \Omega_{000001,111010} &\!=\!& \Omega_{000001,111100} = \Omega_{000010,111100} = b_2,\\
    \Omega_{000011,111011} &\!=\!& \Omega_{000101,111101} = \Omega_{000110,111110} = -b_3,\\
    \Omega_{000011,111101} &\!=\!&  \Omega_{000011,111110} = \Omega_{000101,111110} =
    -b_3.
\end{eqnarray*}
Now an elementary but lengthy calculation shows that the chosen
parameters define a feasible point of the problem and yield a
success probability of $\hbox{tr}\{\Omega\rho_{GHZ}\otimes\id\}=0.8$.

To prove the optimality of this result we now consider the dual
problem. The Lagrange function for the minimization problem in
Eq.\ (\ref{eq6}) is given by
\begin{eqnarray}
    L(\Omega,\lambda_A,\lambda_B,\lambda_C,\lambda_p,\nu)
    &=&\hbox{tr}\{\nu\rho_{GHZ}\otimes(\id-\rho_{W})) \}\nonumber\\
    &&\hspace*{-4cm}+ \hbox{tr}\{\lambda_p\}-\hbox{tr}\{\Omega((\rho_{GHZ}-\lambda_p)\otimes\id +\lambda_A^{\Gamma_A}
    +\lambda_B^{\Gamma_B}+\lambda_C^{\Gamma_C} \},\nonumber
\end{eqnarray}
where
$\lambda_p,\lambda_A^{\Gamma_A},\lambda_B^{\Gamma_B},\lambda_C^{\Gamma_C}\ge
0$. The Lagrange function has to be minimized over all $\Omega\ge
0$. This is feasible only if
\begin{equation}
    \rho_{GHZ}\otimes\id + \lambda_A^{\Gamma_A}
    +\lambda_B^{\Gamma_B}+\lambda_C^{\Gamma_C}-\lambda_p\otimes\id
    -\nu\rho_{GHZ}\otimes(\id-\rho_{W})\le 0 \label{ConstraintB}
\end{equation}
in which case we obtain the dual function
\begin{equation}
    g(\lambda_A,\lambda_B,\lambda_C,\lambda_p,\nu) =
    -\hbox{tr}\{\lambda_p\}.
\end{equation}
Maximizing this function under the constraints
$\lambda_A,\lambda_B,\lambda_C,\lambda_p\ge 0$ and
Eq.\ (\ref{ConstraintB}) yields upper bounds on the success
probabilities of the primal problem. The following trial solution
yields $-tr\{\lambda_p\}=-0.8$ satisfying all the constraints and
matching the value of the primal optimum thereby proving its
optimality. For simplicity we only give the non-zero matrix
elements
\begin{eqnarray*}
    \lambda_{p\,1,1} &=& \lambda_{p\,8,8} = -\lambda_{p\,1,8} =
    -\lambda_{p\,8,1} = 0.4,\\
    \lambda_{A\,1,1} &=& -\lambda_{A\,1,4} = -\lambda_{A\,1,6}
    = \lambda_{A\,4,4} =
    \lambda_{A\,6,6} = \lambda_{A\,8,8}\\
    &=& \lambda_{A\,4,6} = 4\lambda_{A\,5,5} =
    -2\lambda_{A\,5,8} =  0.8/3,\\
    \lambda_{A\,57,57} &=& -\lambda_{A\,57,60} = -\lambda_{A\,57,62} =
    \lambda_{A\,60,60} =
    \lambda_{A\,62,62}\\
    &=& \lambda_{A\,64,64}=\lambda_{A\,60,62} = 4\lambda_{A\,61,61} =
    -2\lambda_{A\,61,64}\\
    &=& 0.8/3,\\
    \nu &=& 1.8.
\end{eqnarray*}
The elements of $\lambda_B$ and $\lambda_C$ are obtained from
$\lambda_A$ by cyclic permutation of the parties $A,B$ and $C$ so
that for example $\lambda_{A\,5,5}=\lambda_{B\,2,2}$. Direct
calculation no shows that this trial solution is feasible for the
dual problem and yields the value $g=-0.8$ which is identical to
that obtained from the trial solution for the primal problem. This
completes the proof of optimality.
\section{From $W$ to GHZ employing non-trace preserving PPT maps}
\label{sec: W to GHZ non-trace preserving}
The optimization of the success probability for the transformation
from $W$ to GHZ proceed along very similar lines as those given in
the previous appendix. Mathematically the problem is formulated as
\begin{eqnarray}
    \hbox{tr}\{\Psi(\rho_{W})\} &=& \hbox{tr}\{\Omega
    \rho_{W}\otimes\id\} \label{wghznewref}
\end{eqnarray}
under the constraints
\begin{eqnarray*}
    \hbox{tr}\{\Omega\rho_{W}\otimes(\id-\rho_{GHZ})\} &\!=\!& 0, \\
        \hbox{tr}_{V'}\{\Omega\} \le \id, \hspace*{0.92cm} \Omega^{\Gamma_{V}}
&\!\ge\!& 0,\\
    (\Omega^{\Gamma_{A}\otimes\Gamma'_{A}})^{\Gamma_{V}}
    \ge  0,\;
    (\Omega^{\Gamma_{B}\otimes\Gamma'_{B}})^{\Gamma_{V}}
    &\!\ge\!&  0,\;
    (\Omega^{\Gamma_{C}\otimes\Gamma'_{C}})^{\Gamma_{V}}
    \ge  0.
\end{eqnarray*}
Symmetries analogous to those presented in the previous sections
hold. Following the arguments analogous to those in section
\ref{sec: Tripartite} most matrix elements of $\Omega$ vanish. In
the following we will present those non-vanishing matrix elements
that are sufficient to reconstruct all the remaining non-zero
elements of the trial solution from the symmetries of the problem.
With $b_1=3b_2/4=3b_3=1/6$
\begin{eqnarray*}
    \Omega_{000001,000001} &\!=\!& \Omega_{111001,111001} = 11/90,\\
    \Omega_{001001,001001} &\!=\!& 4\Omega_{010001,010001} = 4\Omega_{100001,100001} = b_2,\\
    \Omega_{001001,010001} &\!=\!& \Omega_{001001,100001} = -2\Omega_{010001,100001} = -b_2/2,\\
    \Omega_{011001,011001} &\!=\!& \Omega_{101001,101001} = 2\Omega_{110001,110001} =b_1,\\
    \Omega_{011001,101001} &\!=\!& -2\Omega_{011001,110001} = -2\Omega_{101001,110001} = b_3,\\
    \Omega_{000000,000000} &\!=\!& \Omega_{111000,111000} = b_2/2,\\
    \Omega_{001000,001000} &\!=\!& \Omega_{010000,010000} = \Omega_{100000,100000} = b_3,\\
    \Omega_{001000,010000} &\!=\!& \Omega_{001000,100000} = \Omega_{010000,100000} = b_3,\\
    \Omega_{011000,011000} &\!=\!& \Omega_{101000,101000} = \Omega_{110000,110000} = b_3/2,\\
    \Omega_{011000,101000} &\!=\!&  \Omega_{011000,110000} = \Omega_{101000,110000} = b_3/2,\\
    \Omega_{000000,000111} &\!=\!& -7\Omega_{111000,111111}/10 = -7/90,\\
    \Omega_{001000,001111} &\!=\!& \Omega_{010000,010111} = \Omega_{100000,100111} = b_3,\\
    \Omega_{001000,010111} &\!=\!& \Omega_{001000,100111} = \Omega_{010000,100111} = b_3,\\
    \Omega_{011000,011111} &\!=\!& \Omega_{101000,101111} = \Omega_{110000,110111} = b_3/2,\\
    \Omega_{011000,101111} &\!=\!&  \Omega_{011000,110111} = \Omega_{101000,110111} = b_3/2.
\end{eqnarray*}
With this trial solution we find $\hbox{tr}\{\Omega
\rho_{W}\otimes\rho_{GHZ}\}=\frac{1}{3}$.

To prove the optimality of this result we now consider the dual
problem. The Lagrange function for the minimization problem in Eq.
(\ref{wghznewref}) is given by
\begin{eqnarray}
    &&\hspace*{-0.5cm}L(\Omega,\lambda_A,\lambda_B,\lambda_C,\lambda_p,\nu)
    = -\hbox{tr}\lambda_p -\hbox{tr} \!\!\! \sum_{i=A,B,C} \lambda_i^{\Gamma_i\otimes \Gamma'_i}\nonumber\\
    &&-\hbox{tr}\{\Omega( (\rho_W-\lambda_p)\otimes\id
    - \nu \rho_{W}\otimes (\id - \rho_{GHZ})\},
\end{eqnarray}
where $\lambda_p,\lambda_A,\lambda_B,\lambda_C\ge 0$. This
Lagrange function has to be minimized over all $\Omega\ge 0$ which
is feasible only if
\begin{equation}
    (\rho_W-\lambda_p)\otimes\id
    - \nu \rho_{W}\otimes(\id - \rho_{GHZ})
    + \!\!\! \sum_{i=A,B,C} \lambda_i^{\Gamma_i\otimes \Gamma'_i}\le 0,
    \label{constraint}
\end{equation}
in which case we obtain the dual function
\begin{equation}
    g(\lambda_A,\lambda_B,\lambda_C,\lambda_p,\nu) =
    -\hbox{tr}\{\lambda_p\}.
\end{equation}
Now we need to maximize this function under the constraints
$\lambda_A,\lambda_B,\lambda_C,\lambda_p\ge 0$ and Eq.
(\ref{constraint}). Each trial solution gives an upper bound on
the success probability of the primal problem. It turns out that
we can approach the $-\hbox{tr}\{\lambda_p\}=-\frac{1}{3}$ arbitrarily
closely.

We begin by determining all non-zero matrix elements of
$\lambda_p$ in terms of $\lambda_{p\,2,2}$ so that
\begin{eqnarray*}
    \lambda_{p\,3,3} &=& \lambda_{p\,5,5} = \lambda_{p\, 2,2},\\
    \lambda_{p\,2,3} &=& \lambda_{p\,2,5} =
    \lambda_{p\,3,5}=-\lambda_{p\, 2,2}/2.
\end{eqnarray*}
Furthermore, we completely determine the matrices
$\lambda_A,\lambda_B$ and $\lambda_C$. To this end we give all the
nonzero values of $\lambda_A$ as the other matrices are uniquely
determined through cyclic permutations from $\lambda_A$.
\begin{eqnarray*}
    \lambda_{A\,18,23} &=& -\frac{1}{5} = \lambda_{A\,34,39},\\
    \lambda_{A\,18,39} &=& -\frac{3}{10} = \lambda_{A\,34,23}
\end{eqnarray*}
and
\begin{eqnarray*}
    \lambda_{A\,17,17} &=& \lambda_{A\,33,33} = -\lambda_{A\,17,33} =
    -\lambda_{A\,33,17} =  0.1/9,\\[0.2cm]
    \lambda_{A\,18,18} &=& \lambda_{A\,34,34} = 0.3; \;\;
    \lambda_{A\,18,34} = \lambda_{A\,34,18} = 0.2,\\[0.2cm]
    4\lambda_{A\,19,19} &=&  \lambda_{A\,35,35} = 2\lambda_{A\,19,35} =
    2\lambda_{A\,35,19} = 0.4/9,\\[0.2cm]
    \lambda_{A\,20,20} &=& 4\lambda_{A\,36,36} = 2\lambda_{A\,20,36}
    = 2\lambda_{A\,36,20} =0.4/9,\\[0.2cm]
    \lambda_{A\,21,21} &=& 4\lambda_{A\,37,37} = 2\lambda_{A\,21,37} =
    2\lambda_{A\,37,21} = 0.4/9,\\[0.2cm]
    4\lambda_{A\,32,32} &=& \lambda_{A\,38,38} = 2\lambda_{A\,32,38}
    = 2\lambda_{A\,38,32} = 0.4/9,\\[0.2cm]
    \lambda_{A\,23,23} &=& \lambda_{A\,39,39} = 0.3; \;\;
    \lambda_{A\,23,39} = \lambda_{A\,39,23} = 0.2,\\[0.2cm]
    \lambda_{A\,24,24} &=& \lambda_{A\,40,40} = -\lambda_{A\,24,40} =
    -\lambda_{A\,40,24} = 0.1/9.
\end{eqnarray*}
The elements of $\lambda_B$ and $\lambda_C$ are obtained from
$\lambda_A$ by cyclic permutation of the parties $A,B$ and $C$ so
that for example $\lambda_{A\,5,5}=\lambda_{B\,2,2}$. A direct
calculation shows that the constraints
$\lambda_A,\lambda_B,\lambda_C,\lambda_p\ge 0$ are satisfied with
these choices. Now we need to verify whether the constraint
\begin{equation}
    (\rho_W -\lambda_p)\otimes\id - \nu \rho_{W}\otimes(\id -
    \rho_{GHZ}) + \!\!\! \sum_{i=A,B,C}
    \lambda_i^{\Gamma_i\otimes \Gamma'_i}\le 0
    \label{const}
\end{equation}
can be verified as well. Note that we still have the free
parameters $\lambda_{p\, 2,2}$ and $\nu$. A lengthy computation
(preferably employing Mathematica) shows that the left hand side of
the constraint has $6$ distinct nonzero eigenvalues, namely
\begin{eqnarray*}
    && \hspace*{-0.75cm}\mu_1 = 2-\nu,
    \hspace*{2.5cm} \mu_2 = \frac{1}{90}(13-135\lambda_{p\,
    2,2}),\\
    && \hspace*{-0.75cm} \mu_3 = \frac{1}{30}(-2-45\lambda_{p\,
    2,2}),\;\;\;\;\;\; \mu_4 = \frac{1}{30}(4-45\lambda_{p\, 2,2}),\\
    && \hspace*{-0.75cm} \mu_{\pm}=\frac{1}{60}[
    47-30\nu -45\lambda_{p\, 2,2} \pm\nonumber\\[0.1cm]
    &&\hspace*{-0.75cm} \sqrt{1569-2220\nu+3330\lambda_{p\, 2,2}
    +(45\lambda_{p\, 2,2}-30\nu)^2}\;].\label{mupm}
\end{eqnarray*}
Clearly, for $\nu\ge 2$ and $\lambda_{p\, 2,2}\ge\frac{13}{135}$
the first 4 eigenvalues are non-positive. Now we can verify by
direct inspection that for any choice of $\lambda_{p\, 2,2}>1/9$
there is a choice of $\nu>2$ such that the two eigenvalues
$\mu_{\pm}$ are negative so that also the constraint Eq.
(\ref{const}) is satisfied. Therefore, for any value of
$-\hbox{tr}\{\lambda_p\}<-\frac{1}{3}$ we can satisfy the
constraints. This shows that the primal problem which achieves a
success probability $p=1/3$ is optimal.

\section{Single copy distillation from high rank mixed states}
\label{sec: Single copy distillation from high rank mixed states}
In this appendix, we prove that PPT-operations cannot distill any pure
entangled states from a single copy of $\rho$ on
$\mathbb{C}^d\!\otimes\!\mathbb{C}^d$ when $\hbox{rank}(\rho)\!\ge\!d^2\!-\!2$.
To this end, it suffices to show that the success probability
$p(\rho\!\rightarrow\!P^+_{d'})$ under PPT-operations ($\Psi$) in the trace
non-preserving scheme is strictly zero, where $\rho\!\in\!{\cal H}(V)$ and
$P^+_{d'}\!\in\!{\cal H}(V')$. Since both $\rho\otimes P^+_{d'}$ and
$\rho \otimes (\id\!-\!P^+_{d'})$ is invariant under the local unitary
transformation of $\id \otimes \id \otimes U \otimes U^*$, it
suffices to consider $\Omega$ invariant under these local operations, i.e.
\begin{equation}
\Omega=A\otimes P^+_{d'}+B\otimes \frac{\id-P^+_{d'}}{{d'}^2-1},
\end{equation}
with $A$ and $B$ being matrices on ${\cal H}(V)$. The success probability is
then
\begin{equation}
p(\rho\rightarrow P^+_{d'})=\hbox{tr}\{\Omega \rho\otimes P^+_{d'}\}
=\hbox{tr}\{A\rho\},
\end{equation}
and constraints for $\Omega$ are
\begin{eqnarray*}
&& \hbox{tr}\{\Omega \rho\otimes(\id-P^+_{d'})\}=\hbox{tr}\{B\rho\}=0,\\
&& \quad A\ge0, \;\;\; B\ge 0, \;\;\; \id\ge A+B, \\
&& \frac{1}{d'-1}B^{\Gamma_A}\!\ge\! A^{\Gamma_A} \!\ge\! -\frac{1}{d'+1}B^{\Gamma_A}.
\end{eqnarray*}
Since $B\!\ge\!0$ and $\hbox{tr}B\rho\!=\!0$, the support space of $B$
must be contained in the kernel space of $\rho$, and hence
$\hbox{rank}(B)\!\le\!2$ when $\hbox{rank}(\rho)\!\ge\!d^2\!-\!2$.
On the other hand, $B^{\Gamma_A}\!\ge\!0$ must hold from
$\frac{1}{d'-1}B^{\Gamma_A}\!\ge\!-\frac{1}{d'+1}B^{\Gamma_A}$, and $B$ must
be a separable state (leaving out normalization) since
$\hbox{rank}(B)\!\le\!d$ \cite{Horodecki00b}. Therefore, by using appropriate
local basis, $B$ can be written as
\begin{equation}
B=y|11\rangle\langle11|+z|ef\rangle\langle ef|,
\end{equation}
where $y$ and $z$ are non-negative values and
\begin{equation}
|ef\rangle=(\cos u|1\rangle+\sin u|2\rangle)\otimes
(\cos v|1\rangle+\sin v|2\rangle)
\end{equation}
is a product vector. In this choice of local basis, $B^{\Gamma_A}\!=\!B$.
Let $P$ be the projector on the support space of $B^{\Gamma_A}$ and
$Q\!\equiv\!I\!-\!P$. The condition of
$\frac{1}{d'-1}B^{\Gamma_A}\!\ge\!A^{\Gamma_A}\!\ge\!
-\frac{1}{d'+1}B^{\Gamma_A}$
implies that $\pm QA^{\Gamma_A}Q\!\ge\!0$, and hence $QA^{\Gamma_A}Q\!=\!0$
must hold. Furthermore, $A^{\Gamma_A}\!+\!\frac{1}{d'+1}B^{\Gamma_A}$ must be a
positive operator, for which
$Q(A^{\Gamma_A}\!+\!\frac{1}{d'+1}B^{\Gamma_A})Q\!=\!0$ also holds. Therefore,
support space of $A^{\Gamma_A}\!+\!\frac{1}{d'+1}B^{\Gamma_A}$ must be $P$,
and hence the support space of $A^{\Gamma_A}$ must be contained in the support
space of $B^{\Gamma_A}$. As a result,
$\hbox{rank}(A^{\Gamma_A})\!\le\!\hbox{rank}(B^{\Gamma_A})\!\le\!2$.
Furthermore, $A^{\Gamma_A}$ must be written in the form of
\[
A^{\Gamma_A}=r|11\rangle\langle 11|+s|11\rangle\langle ef|
+s^* |ef\rangle\langle 11|+t|ef\rangle\langle ef|,
\]
and $A$ is then given by
\[
A=r|11\rangle\langle 11|+s|e1\rangle\langle 1f|+s^* |1f\rangle\langle e1|
+t|ef\rangle\langle ef|.
\]
Therefore, $A$ must be essentially two-qubit state (leaving out normalization)
since $A\!\ge\!0$ must hold. If the two-qubit state $A$ is entangled,
$\hbox{rank}(A^{\Gamma_A})$ must be 4 \cite{Verstraete01d,Ishizaka04a},
which contradicts that $\hbox{rank}(A^{\Gamma_A})\!\le\!2$. Therefore,
$A$ and $A^{\Gamma_A}$ must be written in a separable form.

In the case where $\sin u \sin v\!\ne\!0$, the support space of $A^{\Gamma_A}$,
which is spanned by $|11\rangle$ and $|ef\rangle$, contains only two product
vectors ($|11\rangle$ and $|ef\rangle$ itself) \cite{Sanpera98a},
and hence $A^{\Gamma_A}$ must be written as
\begin{equation}
A^{\Gamma_A}=r|11\rangle\langle 11|+t|ef\rangle\langle ef|,
\end{equation}
and $A\!=\!A^{\Gamma_A}$. As a result, the support space of $A$ is contained
in the support space of $B$, and hence
$p(\rho\!\rightarrow\!P^+_{d'})\!=\!\hbox{tr}A\rho\!=0$ as
$\!\hbox{tr}B\rho\!=\!0$.
In the case where $\sin u \sin v\!=\!0$, $|e\rangle\!=\!|1\rangle$ or
$|f\rangle\!=\!|1\rangle$ holds. As a result, $A$ is spanned by
$\{|11\rangle,|1f\rangle\}$ (or $\{|11\rangle,|e1\rangle\}$) which is a kernel
of $\rho$, and hence $p(\rho\!\rightarrow\!P^+_{d'})\!=\!\hbox{tr}A\rho\!=\!0$.

%

\begin{thebibliography}{10}
\bibitem{Note_for_PPT_operations}
This is analogous to the fact that every separable operations can be
implemented with smaller but nonzero probability by LOCC (LOCC is supported
by separable states in its own).
However, while the class of separable operations is strictly wider than the class
of LOCC \protect{\cite{Bennett99}}, it is an intriguing open problem whether
the class of PPT operations is strictly wider than the class of LOCC supported
by PPT bound entanglement, or not.
\bibitem{Note_for_unlockable_state}
When $N\!=\!2$ ($AB$) and $N'\!=\!4$ ($CDEF$), for example,
both $(\Omega^{\Gamma_{AC}})^{\Gamma_V}\!\ge\!0$
and
$(\Omega^{\Gamma_{BC}})^{\Gamma_V}\!\ge\!0$
ensure the undistillability with respect to $A$, $B$, and $C$.
Likewise, the undistillability with respect to $D$, $E$, and $F$
is ensured.
\bibitem{Note_for_entanglement_measure}
Let $E_C^{\rm PPT}$ and $E_D^{\rm PPT}$ be the entanglement cost and
distillable entanglement by PPT-operations.
The transformation of $|\phi^+\rangle^{\otimes nE_C^{\rm PPT}(\sigma)}
\!\rightarrow\!\sigma^{\otimes n}\!\rightarrow\!
|\phi^+\rangle^{\otimes nE_D^{\rm PPT}(\sigma)}$ is then possible by
PPT-operations in an asymptotic limit, and hence
a weakly additive and continuous measure ($E$) monotonic under PPT-operations
satisfies $E_C^{\rm PPT}(\sigma)\!\ge\!E(\sigma)\!\ge\!E_D^{\rm PPT}(\sigma)$
(see also \protect{\cite{Horodecki00a}}). The asymptotic relative entropy of
entanglement with respect to PPT-states is an example of such measures.
Moreover, $E_C^{\rm LOCC}\!\ge\!E_C^{\rm PPT}$,
$E_D^{\rm PPT}\!\ge\!E_D^{\rm LOCC}$ and
$E_C^{\rm LOCC}(|\psi\rangle)\!=\!E_D^{\rm LOCC}(|\psi\rangle)$
for a bipartite pure state.
As a result,
$E_C^{\rm PPT}(|\psi\rangle)\!=\!
E(|\psi\rangle)\!=\!\!E_D^{\rm PPT}(|\psi\rangle)$,
and hence every weakly additive continuous measures for bipartite pure states
are equal.
\end{thebibliography}

\end{document}